\documentclass[a4paper]{jpconf}

\usepackage{amsmath,amssymb}
\usepackage{epsfig}
\usepackage[square,sort&compress]{natbib}
\usepackage{graphicx}
\usepackage{float}
\usepackage{array}
\usepackage{rotating}
\usepackage{bm}
\usepackage{url}
\usepackage{color}

\begin{document}
\title{Prospects for gravitational wave astronomy with
next generation large-scale pulsar timing arrays}

\author{Yan Wang$^1$ and Soumya D. Mohanty$^2$}

\address{$^1$ MOE Key Laboratory of Fundamental Physical Quantities Measurements, Hubei Key Laboratory of Gravitation and Quantum Physics,  School of Physics, Huazhong University of Science and Technology, 
1037 Luoyu Road, Wuhan, Hubei Province 430074, China}

\address{$^2$ Department of Physics and Astronomy, The University of Texas Rio Grande Valley, \\
One West University Blvd, Brownsville, TX 78520, USA}

\ead{ywang12@hust.edu.cn; soumya.mohanty@utrgv.edu}

\begin{abstract}


Next generation radio telescopes, namely the Five-hundred-meter 
Aperture Spherical Telescope (FAST) and the Square Kilometer Array (SKA), 
will revolutionize the 
pulsar timing arrays (PTAs) based gravitational wave (GW) searches.
We  review some of the characteristics of  FAST and SKA,
and the resulting PTAs, that are pertinent to the detection of 
gravitational wave signals from individual supermassive 
black hole binaries.

\end{abstract}

\section{Introduction}

The high-frequency ($10-10^{4}$ Hz) gravitational wave (GW) window 
in observational astronomy and cosmology 
has been opened by LIGO with the first direct detection of GWs from the merger 
of a stellar-mass black hole binary \citep{2016PhRvL.116f1102A}. 
Progress on opening up the GW window at lower frequencies 
continues \citep{2017arXiv170200786A, 2016CQGra..33c5010L} with the 
$10^{-9}-10^{-6}$ Hz being probed by pulsar timing arrays (PTAs). 
This frequency regime is expected to host GW signals from 
supermassive black hole binaries residing in the innermost regions of 
merged galaxies. Other potential signals that could be detected are 
stochastic GW background from cosmic strings \citep{2010PhRvD..81j4028O},  
inflation \citep{1979JETPL..30..682S} and 
first-order QCD phase transition \citep{2010PhRvD..82f3511C}. 

Three main regional collaborations are involved in advancing the 
search for GWs with PTAs:
the North American Nanohertz Observatory for Gravitational 
Waves (NANOGrav) \citep{2013ApJ...762...94D} based on 
the Arecibo observatory and Green Bank telescope; 
the Parkes Pulsar Timing Array (PPTA) \citep{2013PASA...30...17M} 
based on the Parkes telescope; the European Pulsar Timing 
Array (EPTA) \citep{2010CQGra..27h4014F} based on five one-hundred 
meter class radio telescopes such as Effelsberg and Lovell. 
These collaborations share observational data and 
analysis tools within the framework of the International Pulsar Timing 
Array (IPTA) \citep{2013CQGra..30v4010M}. The IPTA 
has recently released timing data for 49 millisecond 
pulsars (MSPs) gathered over a period of 5 to 22 
years \citep{2016MNRAS.458.1267V}.

The sensitivities of PTAs to stochastic GW background 
\cite{2015Sci...349.1522S,  2016ApJ...821...13A}, 
continuous waves from individual sources \cite{2014ApJ...794..141A} 
and bursts with memory \citep{2015ApJ...810..150A} 
have constantly improved over the past five years. 
The first PTA-based GW detection, 
possibly a stochastic background signal,
is anticipated in the next a few years. 

Next generation large-scale radio telescopes, such as 
the Five-hundred-meter Aperture Spherical Telescope 
(FAST) \citep{2014arXiv1407.0435H} 
and the Square Kilometer Array (SKA) \citep{2009A&A...493.1161S},
will bring profound changes and challenges in the enterprise of PTA by 
discovering hundreds of new stable MSPs and substantially improving 
timing precision. 


Pulsar timing measures the arrival time of a fiducial phase point (usually the peak) 
on the integrated pulse profile which is obtained by folding the 
individual pulses across integration time and (after dedispersion) 
radio frequency channels.  At 100 ns timing precision level, 
the main contributors in the error budget of the TOA measurement 
are the pulse phase jitter noise (due to the fluctuation in the shape and 
arrival time of individual pulses) and the additive radiometer 
noise (due to sky background and instrumental thermal electron 
noise) that can be estimated respectively by the following 
formulae \citep{2010arXiv1010.3785C, 2015JPhCS.610a2019W}
%
\begin{equation}\label{eq:jitter}
\sigma_{\rm j} \approx 0.28 W\sqrt{\frac{P}{\Delta t}} \;,
\end{equation}
\begin{equation}\label{eq:rad}
\sigma_{\rm r} \approx \frac{W S}{F\sqrt{2\Delta f \Delta t}}\sqrt{\frac{W}{P-W}} \;. 
\end{equation}
%
Here $\Delta t$ is the integration time to obtain an integrated pulse profile, 
$P$ is the rotation period of the pulsar, $W$ is the effective width of the 
integrated pulse profile, $F$ is the flux density, $\Delta f$ is the bandwidth. 
$S=2\eta k_B (A_{\rm eff}/T_{\rm sys})^{-1}$ is the system equivalent 
flux density (SEFD)~\citep{2013tra..book.....W}, 
where $\eta \sim 1.0$ is the system efficiency factor, 
$k_B$ is the Boltzmann's constant, $T_{\rm sys}$ is the system temperature, 
$A_{\rm eff}$ is the effective collecting area, 
and $A_{\rm eff}/T_{\rm sys}$ is called the telescope sensitivity. 
The signal-to-noise ratio (S/N) of integrated pulse profile is 
defined as the height of the pulse profile divided by $\sigma_{r}$.
The total noise of TOA $\sigma_t$  satisfies 
$\sigma_{t}^{2} = \sigma_{j}^{2} + \sigma_{r}^{2}$.


\section{Next generation facilities}

\subsection{FAST}\label{subsec:fast}

FAST is located in the karst depression area in southwest China. It is the  
largest filled-aperture single dish radio telescope in the world,
having a reflector that is $500$~m in diameter with 
$A_{\rm eff}=67,000~{\rm m}^2$ (equivalently an aperture of 
about 300~m diameter).
The active surface and feed cabin suspension systems enable 
a zenith angle coverage of up to $40^{\circ}$ 
which gives a declination range of $66^{\circ}$ to $-14^{\circ}$. 
Potentially, the zenith angle can be extended up to $60^{\circ}$ 
with $A_{\rm eff}=31,000~{\rm m}^2$ 
(equivalently an aperture of 200~m diameter).
First light of FAST was achieved in September 2016 and it is 
currently under testing and commissioning.

Several sets of receivers have been proposed for FAST 
to cover the frequency range from 70 MHz to 
3 GHz \citep{2011IJMPD..20..989N}. 
Among them, a 19-beam L-band receiver with $T_{\rm{sys}}=20$ K 
and bandwidth of $400~\rm{MHz}$ will be used ultimately 
to conduct the pulsar survey and timing. 
Simulations based on pulsar population models and the designed 
system parameters of FAST show that more than 5000 new pulsars 
would be discovered by the 19-beam receiver, among which 
about $10\%$ would be millisecond pulsars. This survey can be carried out 
efficiently with eight hours survey time 
each day for 200 days \citep{2009A&A...505..919S}. 
During the early science run when the active control and 19-beam 
receiver are not available, FAST can discover new pulsars in the drift 
scan mode \citep{2013IAUS..291..577Y,2016RAA....16..151Z}; 
this has already yeilded eight new discoveries so far \cite{CRAFTS}.


\subsection{SKA}\label{subsec:ska} 

SKA is a global effort to build the world's largest radio telescope with 
an effective collecting area of more than one square kilometer 
formed out of two major arrays:  (i) SKA-Mid (350 MHz -- 14 GHz) 
consisting of up to two thousand 15-m parabolic antennas (dishes) 
most of which will be located in South Africa, 
and (ii) SKA-low (50 MHz -- 350 MHz) consisting of up to one million 
dipole antennas located in western Australia. SKA-Mid is the 
relevant instrument for the purpose of high-precision pulsar timing, 
while SKA-Low will help in pulsar detection as well as studying 
the interstellar medium through which radio pulses propagate. 
The SKA can access the entire sky that is visible from 
its geographic latitude to an angle off zenith of at 
least $85^{\circ}$. 

The construction of SKA will be divided into two consecutive 
phases, SKA1 and SKA2. 
The construction of SKA1 is scheduled to last from 2018 to 2023, 
with early science operation starting as soon as 2021. 
Depending on progress in technology demonstration through 
SKA1 and funding, the construction 
of SKA2 is planned to start  in 2023 and finish around 2030. 
Once completed, SKA is intended to be operational for more than fifty 
years, the typical lifetime of major radio telescopes.


In SKA1, 500 stations with each containing about 250 low frequency 
dipole antennas will be constructed for SKA1-Low, 
while SKA-Mid will have about 200 dishes incorporating 
64 already existing ones from one of the SKA precursors, 
MeerKat \cite{2016SPIE.9906E..25B}. 
The sensitivity $A_{\rm{eff}}/T_{\rm{sys}}>2000~\rm{m^2~k^{-1}}$ for SKA1-Low 
and $A_{\rm{eff}}/T_{\rm{sys}}>1000~\rm{m^2~k^{-1}}$ for SKA1-Mid. 
$A_{\rm{eff}}/T_{\rm{sys}}>10^4~\rm{m^2~k^{-1}}$ for SKA2-Mid 
which is mostly relevant for the SKA era PTA.

Study based on astrophysical models of the pulsar population in 
our Galaxy and SKA design parameters shows that SKA1 is 
expected to discover 9000 canonical pulsars and 1500 MSPs 
by a survey with 65 day on SKA1-Low and 130 days on SKA1-Mid. 
For SKA2, up to 14000 canonical 
and  6000 MSPs can be discovered \citep{2009A&A...493.1161S}. 
Due to their high intrinsic rotational stability, combined with the 
improved sensitivity of SKA, a timing uncertainty of  
$< 100$~ns is likely for a substantial fraction of the 
MSPs \cite{2004NewAR..48..993K,2010arXiv1004.3602M}.

\section{Large-scale PTAs}\label{sec:pta}

In the following, we estimate the properties that a PTA can be expected 
to have in the FAST and SKA era. We will focus here on SMBHB sources 
in the frequency band $10^{-9}-10^{-6}$ Hz. 
The Nyquist rate for the 
timing data, corresponding to twice the upper end of this 
frequency range, is one data point every two weeks.

\subsection{Pulsar timing}\label{subsec:time} 


For FAST, Smits et al. \citep{2009A&A...505..919S} found that it 
will take about 24 hours to time all of the 50 brightest MSPs 
once in order to obtain an integrated pulse profile for each pulsar 
at S/N of 500 that is required for high-precision timing. 
This estimate assumes a five minutes of integration time for pulse profile stabilization.

\begin{figure}
\centering
\centerline{\includegraphics[scale=0.55]{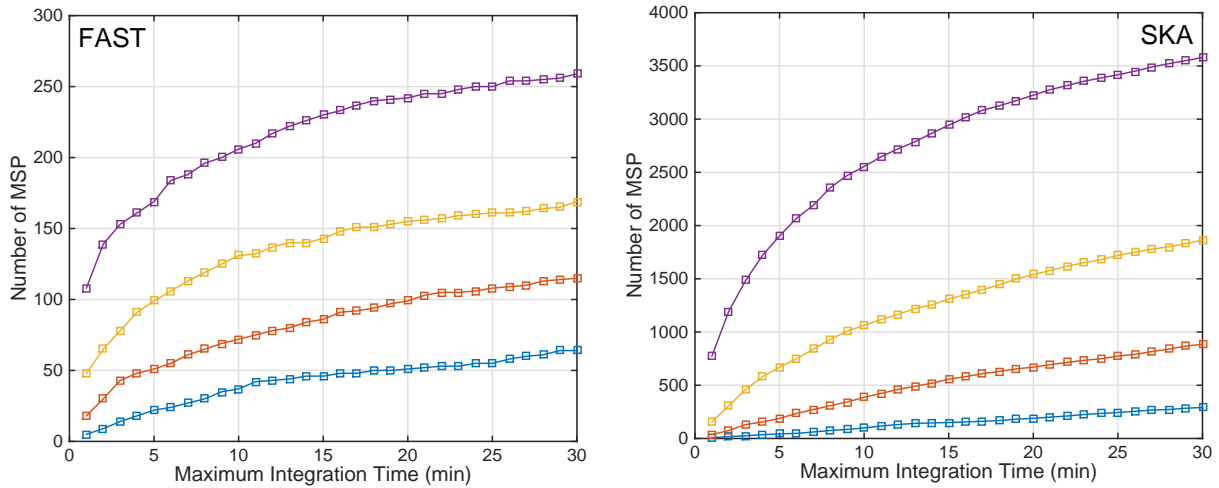}}
\caption{ Number of MSPs having timing noise rms 
$\sigma_t \leq \alpha$ with varying maximum integration time. 
Here $\alpha \in \{50, 100,  200, 500\}$~ns,  
which is indicated by the color lines 
(from bottom to top):  blue (50 ns),  red (100 ns), 
 yellow (200 ns), and purple (500 ns). 
}
\label{fig:fig1}
\end{figure}

It is interesting to see what the make up of a future large-scale PTA 
may look like if the restriction on telescope time allocation for pulsar timing 
were removed. One can then work backwards and deduce how a given 
telescope time allocation will impact the science that can be done. 
In this spirit, we take Eq.~\ref{eq:jitter} and Eq.~\ref{eq:rad} along with 
the design parameters of FAST and SKA to estimate the distribution 
of timing noise rms levels for the MSP population simulated by 
Smits et al.~\cite{2009A&A...505..919S, 2009A&A...493.1161S}. 
Fig.~\ref{fig:fig1} shows the number of MSPs having timing noise rms 
$\sigma_t \leq \alpha$, where $\alpha \in \{50, 100,  200, 500\}$~ns, 
one can expect to see with FAST or SKA given a certain maximum 
integration time per pulsar.  
(Note that this is a gross simplification of the actual situation 
where the integration time will depend on the pulsar brightness.)
Specifically, we see that with FAST, 72 (99, 115) MSPs can be timed to 
be 100 ns or better with 10 (20, 30) mins integration time 
over an observation period of 0.3 (0.6, 1.0) day.

Combining high sensitivity and large field of view 
(FoV, $20~\rm{deg}^2$--$250~\rm{deg}^2$), 
SKA can time MSPs more efficiently in time. 
Smits et al. \citep{2009A&A...493.1161S} found that 
%
%
the observation time for timing 250 MSPs once with 
the best S/N ($\approx 100$) for 
the dense aperture array (AA) configuration will be 6 hours. 
This time can inflate to 15 hours for phased array feed 
15-m dishes and 20 hours for single-pixel feed 15-m dishes. 
%
As before, we can read off from Fig.~\ref{fig:fig1} that, for example, 
189 (387, 557) MSPs can be timed to 
be 100 ns or better with 5 (10, 15) mins integration time, which 
takes 0.7 (1.8, 3.3) day. 
For SKA, the pulse phase jitter noise will dominate the radiometer 
noise for most of bright MSPs due to the high telescope sensitivity. 
In this case, the limiting factor governing timing precision is 
the integration time. 
Therefore, an optimized way to operate telescope, e.g., observing 
multiple MSPs simultaneously \citep{2013CQGra..30v4011L}, 
can be implemented to reduce the aforementioned observation time.

%


\subsection{Single continuous wave source}\label{subsec:cw} 


The non-detection of a stochastic signal from the SMBHB 
population in the most sensitive search carried out till 
date \cite{2015Sci...349.1522S} indicates that SMBHBs 
are sparser than anticipated by population synthesis
models \cite{2009MNRAS.394.2255S}. 
This may increase the likelihood of isolated
SMBHB sources to be the first to be detected by PTAs.
Based on simulations of a cosmological population SMBHBs, 
Rosado et al.~\citep{2015MNRAS.451.2417R} found that 
the probability of detecting single sources within 10 years operation of 
SKA1 will be $40\%$ considering this probability will increase to $80\%$
for SKA2. 
In a different work, Rosado et al.~\citep{2016PhRvL.116j1102R} 
emphasize the role of redshift bias in the detection of massive black 
hole system and how neglecting it underestimates the distance 
reach of current and future PTAs. 
Wang and Mohanty \cite{2017PhRvL.118o1104W} conducted a realistic 
investigation, in the context of individual SMBHB sources, 
of the performance of a SKA-era PTA comprised of 
$10^3$ MSPs selected from the simulated pulsar population 
in  \citep{2009A&A...493.1161S}. 
It was found that a SKA era PTA can significantly increase the 
distance reach to SMBHBs. For example, an all sky search can detect 
systems with redshifted chirp mass of $10^9~M_{\odot}$ to redshift 
$z\sim 1$ and redshifted chirp mass of $10^{10}~M_{\odot}$ to 
redshift $z> 20$.

It is interesting to consider the possibility of merging detections 
of SMBHBs by a SKA-era PTA with ongoing and future searches 
for these systems, through luminosity variations in quasars, in the optical. 
Simultaneous observations in GW and optical can teach us much 
about the accretion physics in SMBHB systems by correlating
the GW signal with optical variability. 
Typical error region for bright GW sources can be expected to 
be $\sim 100~{\rm deg}^2$~\citep{2017PhRvL.118o1104W}, 
which can contain a large number of variable objects. 
However, the frequency of optical variability could be linked 
strongly to the (highly accurate) measured frequency of the 
GW signal and may help in narrowing down the optical counterpart. 
The depth of the search also implies that SMBHB candidates identified 
in the optical, such as PG 1302-102 \cite{2015Natur.518...74G}, 
can be followed up in GWs. A detection of the GW signal from a candidate 
will provide the smoking gun evidence of its true nature.

\subsection{Resolving multiple sources}\label{subsec:resolve}

The focus in current PTA-based GW searches has traditionally 
been on detecting a Gaussian, isotropic stochastic signal. 
However, the true signal is likely to be more complex~\cite{2008MNRAS.390..192S,2011MNRAS.411.1467K}, 
and the multiple source problem~\cite{2012PhRvD..85d4034B} 
is getting more recognition in the PTA community. So far, 
studies~\cite{2012PhRvD..85d4034B,2013PhRvD..87f4036P} of multiple 
source detection have assumed a simplified model in which the signals 
are embedded in white noise, the so-called pulsar term is dropped 
from the waveform of each, and there is no stochastic GW signal.
Under these simplifications, it was shown that the simultaneous fitting of
multiple SMBHB signals can be accomplished using a Genetic Algorithm.
For a realistic assessment of a large-scale PTA, further development 
of data analysis methods that can work without these simplifying
assumptions is required.

The high redshift reach of an SKA era PTA based search for 
SMBHBs implies that multiple resolvable systems may be 
detectable in the PTA data.  
Hence, the data analysis methods for future large-scales PTAs 
should not only be capable of handling $O(10^3)$ pulsars but also 
multiple sources and signal types, ranging from isolated sources 
to the stochastic signal from an unresolved population.

\ack
Y.W. is supported by the National Natural Science Foundation of China 
under grants 11503007, 91636111 and 11690021. 
The contribution of S.D.M. to this paper is partially supported by 
U.S. National Science Foundation grant PHY-1505861. 



\providecommand{\newblock}{}

\end{document}